\documentclass[final]{aipproc}%
\usepackage{amsmath}
\usepackage{amsfonts}
\usepackage{amssymb}
\usepackage{graphicx}
\layoutstyle{8x11double}
\begin{document}
\title{Mesoscopic Thermovoltage Measurement Design}



\author{E.A. Hoffmann}{%
address={Physics Department and Materials Science Institute, University of Oregon,%
Eugene, Oregon 97403-1274, USA},%
altaddress={Physics Department, Ludwig Maximilians University, 80539 Munich, Germany},%
email={eric.hoffmann@physik.lmu.de}
}

\author{H.A. Nilsson}{%
address={Solid State Physics/The Nanometer Structure Consortium, Lund University, Box%
118, S-221 00, Lund, Sweden}%
}
\author{L. Samuelson}{%
address={Solid State Physics/The Nanometer Structure Consortium, Lund University, Box%
118, S-221 00, Lund, Sweden}%
}

\author{H. Linke}{%
address={Physics Department and Materials Science Institute, University of Oregon,%
Eugene, Oregon 97403-1274, USA},%
altaddress={Solid State Physics/The Nanometer Structure Consortium, Lund University, Box%
118, S-221 00, Lund, Sweden},%
email={heiner.linke@ftf.lth.se}%
}

\begin{abstract}
Quantitative thermoelectric measurements in the mesoscopic regime require
accurate knowledge of temperature, thermovoltage, and device energy
scales. We consider the effect of a finite load resistance on thermovoltage
measurements of InAs/InP heterostructure nanowires. Load resistance and ac
attenuation distort the measured thermovoltage therefore complicating the
evaluation of device performance. Understanding these effects improves
experimental design and data interpretation.
\end{abstract}

\maketitle

Touted for their ability to optimize electronic properties while
simultaneously reducing phononic losses, nanostructure materials have advanced
the thermoelectric performance benchmark beyond what is possible with bulk
materials. As a result, many of the underlying mesoscopic processes
responsible for the observed improvements in thermoelectric performance--such
as phonon scattering \cite{Hochbaum08}, phonon drag \cite{Heath08,
Chickering09}, and electron energy filtering \cite{Humphrey02, Datta09}--are
now popular topics in fundamental physics research. The goal of this research
is to provide quantitative understanding of mesoscopic effects and ultimately
improve thermoelectric device design. From an electronic perspective, a
quantum dot (QD) is an appealing test platform due to its strong energy
modulation and tunability. Here we discuss the influence of a finite load
resistance in series with a thermally biased QD defined in an InAs
nanowire.\begin{figure}[tbh]
\includegraphics{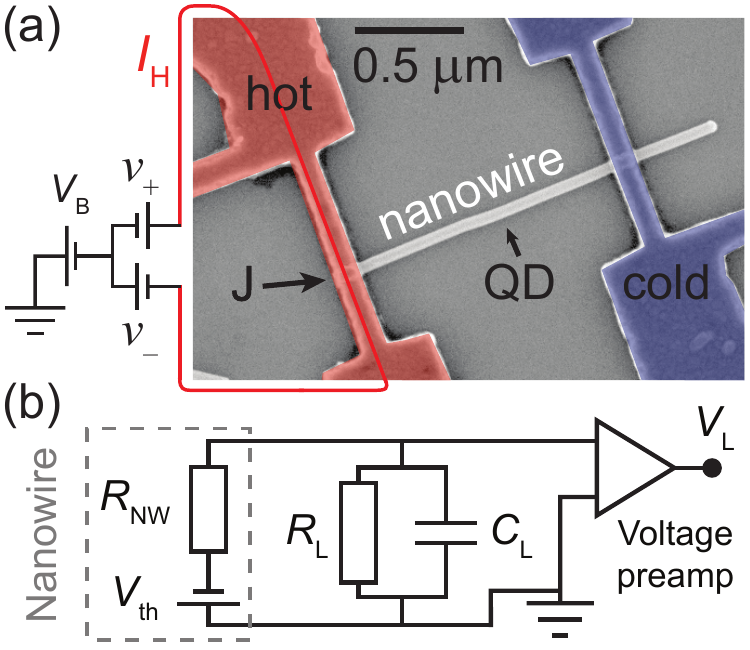}\caption{(a) An SEM image of a nanowire contacted to
hot and cold metallic contacts. The embedded quantum dot (QD) is not resolved
at this resolution. Two ac voltages $v_{\pm}$ create an electrical heating
current, $I_{\text{H}}$, which heats the local electron gas via Joule heating.
$v_{\pm}$ are balanced so that $v_{+}+v_{-}=0$ at the nanowire-contact
junction, $J$. Therefore, the nanowire is electrically biased by $V_{\text{B}%
}$ only. (b) A wiring diagram of the setup during voltage measurements. The
heated nanowire behaves like a battery at voltage $V_{\text{th}}$ with an
internal resistance $R_{\text{NW}}$. The voltage preamplifier measures the
voltage, $V_{\text{L}}$, across an external load resistance, $R_{\text{L}}$,
and load capacitance, $C_{\text{L}}$.}%
\label{device}%
\end{figure}\begin{figure}[th]
\includegraphics{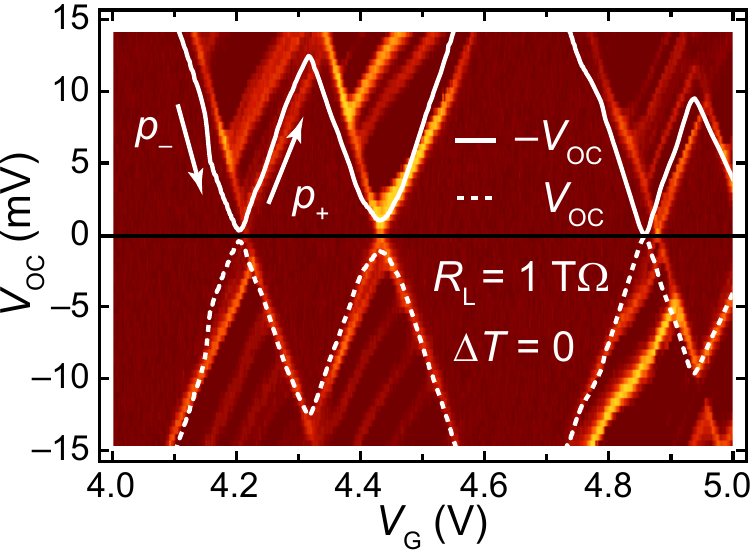}\caption{The open-circuit voltage,
$V_{\text{OC}}$, and its reflection, $-V_{\text{OC}}$, versus gate voltage,
$V_{\text{G}}$, are overlaid on a Coulomb-blockade diagram that was measured
separately. $V_{\text{OC}}$ drifts up and down while passing through QD
resonances generating slopes $p_{\pm}=dV_{\text{OC}}/dV_{\text{G}}$ that match
the Coulomb-blockade diamonds.}%
\label{Voc on Diamond}%
\end{figure}

The experimental device is an InAs/InP heterostructure nanowire grown using
chemical beam epitaxy and is roughly 50 nm in diameter and 1 $\mu$m long (see
Fig.~\ref{device}a). The InAs quantum dot is created by introducing two InP
barriers during the growth process. The nanowire rests on a wafer with a 100
nm insulating SiO$_{\text{x}}$ capping layer above n-doped Si, which operates
as a global back-gate. Ohmic Ni/Au electrical contacts are defined at both
ends of the nanowire using electron beam lithography. The temperature
difference across the quantum dot, $\Delta T$, is provided by Joule heating
generated by an ac heating current at frequency $\omega$ (see
Fig.~\ref{device}a). Further details regarding the application and detection
of $\Delta T$ can be found elsewhere \cite{HoffmannNL, HoffmannJLTP}.
Thermovoltage is detected using a voltage preamplifier by measuring either the
frequency-doubled $2\omega$ signal with a lock-in amplifier or by measuring a
time-averaged dc signal, see Fig.~\ref{device}b. The lock-in technique
provides a better signal-to-noise ratio than the dc technique, but suffers
from $RC$ attenuation owing to the large resistances in the system and the
finite load capacitance, $C_{\text{L}}$, that exists in the cryostat wiring
and filtering. For example, with an off-resonance nanowire resistance of
$R_{\text{NW}}=50$ M$\Omega$ and $C_{\text{L}}=0.5$ nF, which is typical for
our setup, the cutoff frequency is $f_{\text{C}}=1/2\pi R_{\text{NW}%
}C_{\text{L}}=6$ Hz. Therefore, we measure dc thermovoltage to avoid
attenuation effects.\begin{figure}[tbh]
\includegraphics{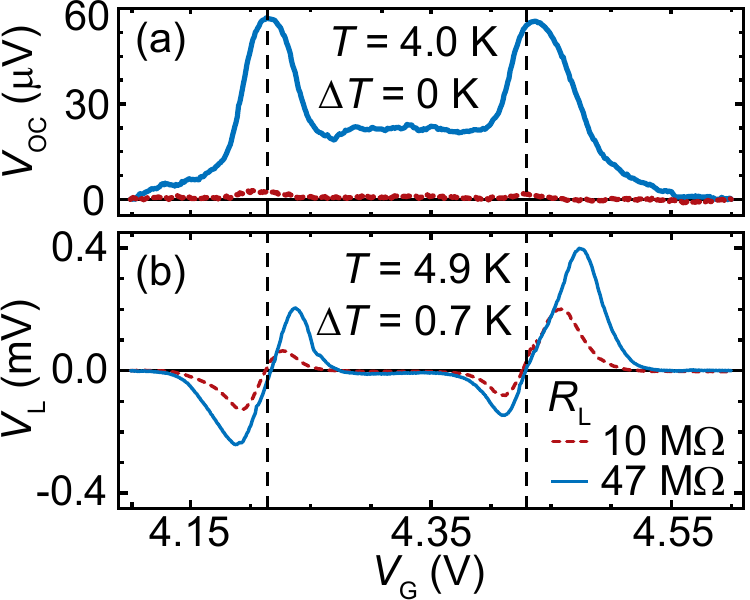}\caption{a) $V_{\text{OC}}$ and b) $V_{\text{L}}$
measured versus $V_{\text{G}}$ using two different $R_{\text{L}}$ values.
$V_{\text{G}}$ spans two QD resonances indicated by the vertical dashed lines.
Decreasing $R_{\text{L}}$ drastically decreases $V_{\text{OC}}$ (compare to
Fig.~\ref{Voc on Diamond}). $V_{\text{OC}}$ is subtracted from subsequence
$V_{\text{L}}$ measurements at finite $\Delta T$.}%
\label{Vl}%
\end{figure}

When measuring thermovoltage, the hot electrochemical potential is grounded
(by the heating circuit) so that $\mu_{\text{H}}=0$ while the cold
electrochemical potential, $\mu_{\text{C}}$, is allowed to float arbitrarily
so that the nanowire can find its equilibrium thermovoltage, $eV_{\text{th}%
}=\mu_{\text{H}}-\mu_{\text{C}}$. Ideally, $V_{\text{th}}$ is measured using a
voltage preamplifier (FEMTO DLPVA-100-F series) with a large input impedance
of 1 T$\Omega$, which separates $\mu_{\text{C}}$ from the preamplifier.
However, the global back-gate, at voltage $V_{\text{G}}$, complicates the
measurement. $V_{\text{G}}$ and $\mu_{\text{C}}$ are capacitively coupled, and
therefore $V_{\text{G}}$ influences $V_{\text{th}}$. For example, when
measuring the \textit{isothermal} open-circuit voltage, $V_{\text{OC}}=\left.
V_{\text{th}}\right\vert _{\Delta T=0}$, suppose $\mu_{\text{C}}$ is in
resonance with a QD level at energy $E\left(  N\right)  $. When $V_{\text{G}}$
increases, $\mu_{\text{C}}$ decreases (as do the QD energy levels) until
$E\left(  N+1\right)  =\mu_{\text{H}}$, at which point $\mu_{\text{C}}$ begins
to increase as electrons flow through the QD from $\mu_{\text{H}}$ to
$\mu_{\text{C}}$ via the $\left(  N+1\right)  $th energy level. This
phenomenon is shown in Fig.~\ref{Voc on Diamond} where, on top of a
Coulomb-blockade stability diagram, $\pm V_{\text{OC}}$ ($-V_{\text{OC}}$ is
the reflection of $V_{\text{OC}}$ across the zero bias line) are shown versus
$V_{\text{G}}$. In fact, $\pm V_{\text{OC}}$ follow the Coulomb-blockade
diamonds of the QD. Using the constant interaction model \cite{NATO}, a
theoretical study of $V_{\text{OC}}$ versus $V_{\text{G}}$ supports our
experimental result by predicting that the slopes $p_{\pm}=dV_{\text{OC}%
}/dV_{\text{G}}$ are identical to the diamond slopes obtained in standard
Coulomb-blockade conductance spectroscopy.

In order to measure, $V_{\text{th}}$, the parasitic $V_{\text{OC}}$ is reduced
by adding an external load resistance, $R_{\text{L}}$, to the system (see Fig.
\ref{device}b). Current flows to ground through $R_{\text{L}}$ and dissipates
$V_{\text{OC}}$, but unfortunately also dissipates $V_{\text{th}}$.
Specifically, when $R_{\text{L}}$ is finite, current flows through the system,
and the preamp measures the induced voltage $V_{\text{L}}$ across the load
resistor. $V_{\text{L}}\neq V_{\text{th}}$ because $V_{\text{th}}$ is (by
definition) the thermally induced voltage at zero current. Simple circuit
analysis reveals that (dc) $V_{\text{th}}=\phi V_{\text{L}}$, where
$\phi=1+R_{\text{NW}}/R_{\text{L}}$ is a function of $V_{\text{G}}$ as given
by the nanowire conductance, $G_{\text{NW}}=1/R_{\text{NW}}$. Therefore, when
choosing $R_{\text{L}}$, a tradeoff must be made between decreasing
$V_{\text{OC}}$ while maintaining $V_{\text{th}}\approx V_{\text{L}}$. We have
best results when $R_{\text{L}}$ is between the on- and off-resonance values
of $R_{\text{NW}}$, that is, when $R_{\text{L}}\approx\left(  \min
R_{\text{NW}}+\max R_{\text{NW}}\right)  /2$. As shown in Fig.~\ref{Vl},
$V_{\text{OC}}$ $\ll V_{\text{L}}$ at $\Delta T=0.7$ K when $R_{\text{L}}=47$
M$\Omega$, and in comparison, this $V_{\text{L}}$ is roughly twice as large as
its $R_{\text{L}}=10$ M$\Omega$ version at finite $\Delta T$.

This research was supported by The Nanometer Structure Consortium at Lund
University, ONR, Energimyndigheten, the Foundation for Strategic Research
(SSF), the Knut and Alice Wallenberg Foundation, and an NSF-IGERT Fellowship.





\bibliographystyle{aipproc}

\end{document}